\documentclass[showpacs,preprintnumbers,
amsmath,amssymb,aps,prd,nofootinbib,superscriptaddress,citeeqsecnum,notitlepage,11pt]{revtex4-1}
\usepackage[T1]{fontenc}
\usepackage{amsmath,xcolor}
\usepackage{amsmath, amssymb,ascmac,fancybox,mathrsfs}
\usepackage{color,booktabs,fancyhdr,bm,comment,cancel}
\usepackage{hyperref}
\usepackage{graphicx,graphics}

\makeatletter
\renewcommand{\p@subsection}{}
\makeatother

\makeatletter
\newcommand{\xequal}[2][]{\ext@arrow 0055{\equalfill@}{#1}{#2}}
\def\equalfill@{\arrowfill@\Relbar\Relbar\Relbar}
\makeatother

\makeatletter

\@addtoreset{equation}{section} 
\makeatother

\newcommand{\chushi}[1]{ }

\newcommand{\Tr}{ \mathrm{Tr} }

\newcommand{\angleLR}[1]{ \left \langle #1 \right \rangle }
\newcommand{\angleN}[1]{ \langle #1 \rangle }

\newcommand{\roundLR}[1]{ \left( #1 \right) }
\newcommand{\roundN}[1]{ ( #1 ) }

\newcommand{\roundB}[1]{ \biggl( #1 \biggr) }

\newcommand{\squareLR}[1]{ \left[ #1 \right] }
\newcommand{\squareN}[1]{ [ #1 ] }

\newcommand{\squareB}[1]{ \biggl[ #1 \biggr] }

\newcommand{\absLR}[1]{ \left| #1 \right| }
\newcommand{\absN}[1]{ | #1 | }

\newcommand{\aabsN}[1]{ \| #1 \| }

\newcommand{\bra}[1]{ \left \langle #1 \right| }
\newcommand{\ket}[1]{ \left| #1 \right \rangle }

\newcommand{\braN}[1]{ \langle #1 | }
\newcommand{\ketN}[1]{ | #1 \rangle }

\let\calccommentout\iffalse 
\let\calcshow\iftrue

\newcommand{\eq}[1]{\begin{equation}\begin{split} #1 \end{split}\end{equation}}

\newcommand {\mathsym}[1]{{}}
\newcommand {\unicode}[1]{{}}

\begin{document}
\count\footins = 1000

\title [mode = title]{Background magnetic field and quantum correlations in the Schwinger effect }                      

\author{Sourav Bhattacharya}

\author{Shankhadeep Chakrabortty}

\author{Hironori Hoshino}

\author{Shagun Kaushal}

\affiliation{Department of Physics, Indian Institute of Technology Ropar,
Rupnagar, Punjab 140 001, India}

\begin{abstract}
\noindent
In this work we consider two complex scalar fields distinguished by their masses coupled to constant background electric and magnetic fields in the $(3+1)$-dimensional Minkowski spacetime and subsequently investigate a few measures quantifying the quantum correlations between the created particle-antiparticle Schwinger pairs. Since the background magnetic field itself cannot cause the decay of the Minkowski vacuum, our chief motivation here is to investigate the interplay between the effects due to the electric and magnetic fields. We start by computing the entanglement entropy for the vacuum state of a single scalar field.
Second, we consider some maximally entangled states for the two-scalar field system  and compute the logarithmic negativity and the mutual information. Qualitative differences of these results pertaining to the charge content of the states are emphasised. Based upon these results, we suggest some possible  effects of a background  magnetic field on the  degradation of entanglement between states in an accelerated frame, for charged quantum fields.  
\end{abstract}

\maketitle
\section{Introduction}
\noindent
Correlation between the states or entanglement is one of  the fundamental characteristics of quantum mechanics.
There are several measures quantifying such correlations, studied in a wide range of theoretical researches, e.g.~\cite{Plenio:2007zz, Horodecki:2009zz, Adesso:2016ygq, Werner:1989zz, Zyczkowski:1998yd, Vidal:1998re, Vidal:2002zz, Plenio:2005, Calabrese:2012nk, Nishioka:2018khk} and references therein.
These correlations constitute the foundation of quantum information theory, see~\cite{NielsenChuang} and references therein.

A natural framework to study  quantum entanglement is a system where pair creation can take place. This includes, most popularly, spacetimes endowed with non-extremal Killing horizons, e.g.~\cite{ Alsing:2003es, FuentesSchuller:2004xp, MartinMartinez:2010ar} or the cosmological scenario, e.g.~\cite{Fuentes:2010dt, Kanno:2014bma, Vennin, Maldacena:2015bha, dePutter:2019xxv, Bhattacharya:2019zno, Bhattacharya:2018yhm, Matsumura:2020uyg} (also references therein). We also refer our reader to e.g.~\cite{Ryu:2006bv, Ryu:2006ef, Calabrese:2012ew, Calabrese:2014yza, Rangamani:2014ywa} and references therein for discussions on quantum entanglement from the holographic perspective.

In this work, we wish to investigate some measures of quantum correlations (namely, the vacuum entanglement entropy, the logarithmic negativity and mutual information for entangled states) in the context of the Schwinger pair creation mechanism~\cite{Schwinger:1951nm, Parker:2009uva}. The entanglement entropy and some other correlation measures for pairwise modes for such a system with a background electric field was studied in~\cite{Ebadi:2014ufa, Li:2016zyv}.  See also~\cite{Dai:2019nzv, Xia:2019ztf, Gavrilov:2019vyi, Li:2018twv} for subsequent developments. 

It is well known that in quantum electrodynamics  a magnetic field itself cannot give rise to pair creation but can affect its rate  if a background electric field is also present, see e.g.~\cite{Karabali:2019oxq} and references therein (see  also~\cite{Karabali:2019ucc}. For discussions on  non-Abelian gauge theory and \cite{Agarwal:2016cir} and references therein for discussions on the notion of entanglement with a quantised gauge filed).  Thus it seems interesting to ask: what will be the effect of a background magnetic field on the quantum correlations between the particle-antiparticle pairs? We may intuitively expect {\it a priori} that the magnetic field will oppose the effect of the electric field. However, how do these correlations explicitly depend upon the magnetic field strength, e.g., are they monotonic?
How do these behaviour differ subject to the charge content of the state we choose?
We wish to address these questions in this work for a complex scalar field in the Minkowski spacetime in $(3+1)$-dimensions. 

The rest of the  paper is organised as follows.
We review very briefly the relevant information quantities in Section \ref{sec:IQ} for the convenience of reader and obtain the solution of the complex scalar's mode functions with the background electromagnetic field in Section \ref{sec:ComplexScalar}.
We compute the vacuum entanglement entropy for a single scalar field in Section \ref{sec:vacuum}, and the logarithmic negativity and mutual information for maximally entangled states of the two scalar fields in Section \ref{sec:bipartite}. We emphasise the qualitative differences of the results subject to the   charge content  of the states.
Finally, we summarise and discuss our results and related issues in Section \ref{sec:SD}. In particular, we speculate  that the well known degradation of the quantum entanglement in an accelerated frame~e.g.~\cite{MartinMartinez:2010ar}, can perhaps be restored for a charged field, upon application of a `strong enough' magnetic field.

We work with the mostly positive signature of the metric and set $\hbar = c = 1$ throughout. The logarithms are understood as $\log_2$ in our numerical calculations.

\section{Measures of correlations -- a quick look}
\label{sec:IQ}
\noindent
Following e.g.~\cite{Plenio:2007zz},
let us consider a bipartite system constituted by subsystems, $A$ and $B$, so that the 
 Hilbert space can be decomposed  as $\mathcal{H}_{AB} = \mathcal{H}_{{A}} \otimes \mathcal{H}_{{B}}$.
Let $\rho_{AB}$ be the density matrix of states on $\mathcal{H}_{AB}$. The reduced density matrix operator of the subsystem $A$ is defined by 
$
\rho_A
=
	\Tr_B
	\rho_{AB}
$,
where the partial traces $\Tr_B$ is taken only over the Hilbert space $\mathcal{H}_B$.

\subsection{Entanglement entropy}
\label{sec:EE}
\noindent
The entanglement entropy of $A$ is defined as the von Neumann entropy of $\rho_A$:
$
	S(\rho_A)
=
	-
	\Tr_A
	\left(\rho_A
	\log \rho_A\right)
$.
When $\rho_{AB}$ corresponds to a pure state, one has $S(\rho_A)=S(\rho_B)$, and it is zero when $\rho_{AB}$ is also separable.
The von Neumann entropies satisfy a subadditivity: $S(\rho_{AB}) \leq 	S(\rho_A) + S(\rho_B)$, where $S(\rho_{AB})$ is the Von Neumann entropy of $\rho_{AB}$.
The equality holds if and only if $\rho_{AB}=\rho_A \otimes \rho_B  $.
More details on these properties can be found in e.g.~\cite{NielsenChuang}.

\subsection{Quantum mutual information}
\label{sec:QMI}
\noindent
The quantum mutual information is a measure of  quantum as well as classical correlations between the subsystems $A$ and $B$. 
For the state $\rho_{AB}$, it is defined as
$
    I(A,B)
=
	S(\rho_{A})
	+
	S(\rho_{B})
	-
	S(\rho_{AB})
$.
The lower bound of the mutual information, $I(A,B) \geq 0$, is immediately obtained by the subadditivity of the entanglement entropy, where
the equality holds  only if $\rho_{AB} = \rho_A \otimes \rho_B$.
Further properties of the quantum mutual information can be found in e.g.~\cite{NielsenChuang}.

\subsection{Entanglement negativity and logarithmic negativity}
\noindent
Even for mixed states, there is a measure of the entanglement of bipartite states~\cite{Zyczkowski:1998yd,Vidal:2002zz}, called the entanglement negativity, defined as
$
    \mathcal{N}(\rho_{AB})
=
	\frac{1}{2}
	\roundN{
	\aabsN{\rho_{AB}^{\text{T}_A}}_1
	-
	1
	}
$, 
where $\rho_{AB}^{\text{T}_A}$ is the partial transpose of $\rho_{AB}$ with respect to the subspace of $A$, i.e., $\roundN{ \ketN{i}_{\! A} \hspace{-0.2ex} \braN{n} \otimes \ketN{j}_{\! B} \hspace{-0.2ex} \braN{\ell} }^{\text{T}_A}: = \ketN{n}_{\! A} \hspace{-0.2ex} \braN{i} \otimes \ketN{j}_{\! B} \hspace{-0.2ex} \braN{\ell}$.
Here, $\aabsN{\rho_{AB}^{T_A}}_1$ is the trace norm, $\aabsN{\rho_{AB}^{T_A}}_1= \sum_{i=1}^{\text{all}} \absLR{\mu_i}$, where $\mu_i$ is the $i$-th eigenvalue of $\rho_{AB}^{T_A}$. 
The logarithm of $\aabsN{\rho_{AB}^{T_A}}_1$ is called the logarithmic negativity, which can be written as $L_{\mathcal{N}} (\rho_{AB})
=
	\log \roundLR{ 1 + 2 \mathcal{N}(\rho_{AB}) }$.
These quantities are entanglement monotones which do not increase under local operations and classical communications.

These quantities measure violation of the positive partial transpose (PPT) in $\rho_{AB}$.
The PPT criterion can be stated as follows. If $\rho_{AB}$ is separable, the eigenvalues of $\rho_{AB}^{T_A}$ are non-negative.
Hence, if $\mathcal{N} \not = 0$ ($L_\mathcal{N} \not = 0$), $\rho_{AB}$ is an entangled state.
On the other hand, if $\mathcal{N} = 0$ ($L_\mathcal{N} = 0$), we cannot judge the existence of the entanglement from this measure, since there exist PPT and entangled states in general.
However, the logarithmic negativity can be useful since it is a calculable measure. Further discussions on it can be found in e.g.~\cite{Horodecki:2009zz}.

\section{Complex scalar in background electromagnetic field}
\label{sec:ComplexScalar}
\noindent
Let us now focus on the complex scalar field theory coupled to external or background electromagnetic fields in the four-dimensional Minkowski spacetime, where the presence of electric field creates particle-antiparticle pair and magnetic opposes this phenomenon.
Our analysis in this section is in parallel with~\cite{Ebadi:2014ufa, Gabriel:1999yz, Bavarsad:2017oyv}.

The Klein-Gordon equation reads  
\eq{
	\roundLR{
	D_\mu D^\mu - m^2 
	}
	\phi(t,\vec x)
&=
	0
,
\label{eq:KGeq}
}
where  $D_\mu = \partial_\mu - i q A_\mu$ is the gauge covariant derivative and $q$ stands for the electric charge of the field.
We consider the external gauge field as $A_\mu = (Ez,-By,0,0)$, where the electric field $E$ and the magnetic field $B$ are constants.

We quantise the field as,
\eq{
	\phi(x)
&=
	\sum_{n_L}
	\int
	\frac{dk^0 dk^x}{\sqrt{4\pi k^0}}
	\squareB{
	a_{k,n_L}
	\phi^{(+)}_{k,n_L} 
    +
	b_{k,n_L}^\dagger
	\left(
	\phi^{(-)}_{k,n_L} 
	\right)^*
	}
,
\label{eq:phi}
}
where $k^0$ is restricted to be positive, and $a_{k,n_L}$ ($b_{k,n_L}^\dagger$) corresponds to the annihilation (creation) operator for the particle (antiparticle).
$n_L=0,1,2, \dots$ stands for the Landau level.
To shorten the notation, we just suppress the label $n_L$. 
The mode functions $\phi^{(\pm)}_{k} $ are given by
\eq{
	\phi^{(+)}_{k} 
=
    e^{-i(k^0t-k^x x)}
	\phi^{(p)}_{k} (y,z)
, \qquad \qquad 
	\left(
	\phi^{(-)}_{k} 
	\right)^*
=
	e^{i(k^0t-k^x x)}
	\left(
	\phi^{(a)}_{k} (y,z)
	\right)^*
,
}
where $p (a)$ stands for particle (antiparticle). Eq.~(\ref{eq:KGeq}) gives,
\eq{
	\squareLR{
	\roundLR{
	k^0 + qEz
	}^2
	-
	\roundLR{
	k^x + qBy
	}^2
	+
	\partial_y^2
	+
	\partial_z^2
	-
	m^2
	}
	\phi^{(\pm)}_{k} (y,z)
&=
	0
.
\label{eq:EOM}
}
We consider a particle that is incoming in the $z$-direction at $\absN{z}\to\infty$.
The independent solutions of (\ref{eq:EOM}) with this boundary condition is derived as
\eq{
	\phi^{(p) \text{in} }_{k} (y,z)
&=
	N^{-1}
	e^{-{y}_+^2/2}
	H_{n_L}({y}_+)
	D_\nu (\zeta_+)
,
\qquad\qquad
	\squareLR{
	\phi^{(a) \text{in}}_{k} (y,z)
	}^*
=
	N^{-1}
	e^{-{y}_-^2/2}
	H_{n_L}({y}_-)
	\squareLR{
	D_\nu (\zeta_-)
	}^*
,
}
where $H_{n_L}({y}_\pm)$ is the Hermite polynomial, and $D_\nu (\zeta_\pm)$ is the parabolic cylinder function.
The variables ${y}_\pm$ and $\zeta_\pm$ are defined by
$
	{y}_\pm
=
	\sqrt{\absLR{qB}}
	\roundN{y \pm {k^x}/{qB}}
$ and
$
	\zeta_\pm
=
	e^{i \pi/4}
	\sqrt{2\absLR{qE}}
	\roundN{
	z \pm {k^0}/{qE}
	}
$, respectively.
Also,
$
	\nu
=
	-(1+i\mu)/2
$,
with the parameter $\mu$ given by
\begin{eqnarray}
	\mu
=
	\frac{ m^2+\absLR{qB}(2n_L+1) }{\absLR{qE}}.
\label{mu}
\end{eqnarray}
From now on we consider that the variation of $\mu$ is solely dependent on the magnetic field $B$,  keeping all the other parameters at some {\it fixed values}. 
In particular, zero value of $\mu$  below will approximately imply vanishing magnetic field, along with the restriction $m^2/|q E| \ll 1$, which can be achieved by applying a `sufficiently strong' electric field.

The incoming modes, $\phi^{(\pm) \text{in}}_{k} (x)$, satisfy the orthonormality conditions, 
defined via the Klein-Gordon inner product, $\angleN{\phi_1, \phi_2} = i \int d^3\vec x \roundN{\phi_1^* \partial_t \phi_2 - \phi_2 \partial_t \phi_1^*}$. Using the properties of the parabolic cylinder functions~\cite{Erdelyi:1953}, we can straightforwardly check that
\eq{
  \angleLR{
    \phi^{(+) \text{in}}_{k} (x)
    ,
    \phi^{(+) \text{in}}_{k'} (x)
    }
&=
    -
    \angleLR{
    \left(
    \phi^{(-) \text{in}}_{k} (x)
    \right)^*
    ,
   \left(
    \phi^{(-) \text{in}}_{k'} (x)
    \right)^*
    }
=
    \delta(k^0 - k^0{}')
    \delta(k^x - k^x{}')
    \delta_{n_L n_L'}
,
\\
    \angleLR{
    \phi^{(+) \text{in}}_{k} (x)
    ,
    \left(
    \phi^{(-) \text{in}}_{k'} (x)
    \right)^*
    }
&=
    0.
\label{eq:Orthonormal}
}

Similarly, we find the orthonormal outgoing modes for particles 
$
	\phi^{(p) \text{out}}_{k} 
\propto
	e^{-{y}_+^2/2}
	H_{n_L}({y}_+)
	\squareN{
	D_\nu(-\zeta_+)
	}^*
$ and for antiparticles
$
	\squareN{
	\phi^{(a) \text{out}}_{k} 
	}^*
\propto
	e^{-{y}_-^2/2}
	H_{n_L}({y}_-)
	D_\nu(-\zeta_-)
$.
These modes also satisfy the orthonormality conditions in the same way as (\ref{eq:Orthonormal}).

The incoming and the outgoing modes furnish two independent quantisations of the scalar field. These modes are related via the Bogoliubov transformation,
\eq{
	\phi^{(+) \text{in} }_{k} 
&=
	\alpha_k
	\phi^{(+) \text{out} }_{k} 
	+
	\beta_k
	\left(
	\phi^{(-) \text{out} }_{-k} 
	\right)^*
,
\label{eq:Bogoliuboftrf0}
}
where $\alpha_k$ and $\beta_k$ are the Bogoliubov coefficients.
The relation (\ref{eq:Bogoliuboftrf0}) yields,
\eq{
	D_\nu(\zeta_+)
&=
	\alpha_k
	\squareLR{
	D_\nu(-\zeta_+)
	}^*
	+
	\beta_k
	D_\nu(-\zeta_+)
,
}
where $
	\squareLR{
	D_\nu(-\zeta)
	}^*
=
	D_{-\nu-1}(i\zeta)
$.
Using the relation \cite{Erdelyi:1953}, 
\eq{
	D_\nu(\zeta)
&=
	e^{-i\pi \nu}
	D_\nu(-\zeta)
	+
	\frac{\sqrt{2\pi}}{\Gamma(-\nu)}
	e^{-\frac{i\pi (\nu+1)}{2}}
	D_{-\nu-1}(i\zeta)
,
}
we obtain
\eq{
	\alpha_k
&=
	\frac{\sqrt{2\pi}}{\Gamma(-\nu)}
	e^{-\frac{i\pi (\nu+1)}{2}}
,
\qquad
	\beta_k
=
	e^{-i\pi \nu}
,
\label{eq:defofalphabeta}
}
which satisfy $\absN{\alpha_{k}}^2 - \absN{\beta_{k}}^2=1$.
Employing the orthonormality conditions, we derive the transformations for the creation and annihilation operator as
\eq{
	a^{\text{in}}_k
=
	\alpha_k
	a^{\text{out}}_k
	-
	\beta_{k}
	b^{\text{out} \dagger}_{-k},
\qquad \qquad 
	b^{\text{in}}_{-k}
=
	-
	\beta_{k}
	a^{\text{out} }_{k}{}^\dagger
	+
	\alpha_{k}
	b^{\text{out}}_{-k}
.
\label{eq:Bogoliuboftrf}
}
Being equipped with this, we are now ready to investigate the correlation properties.

\section{Entanglement entropy for the vacuum }
\label{sec:vacuum}
\noindent
We consider first the vacuum state of incoming modes.
The Hilbert space $\mathcal{H}$ is constructed by the tensor product, $\mathcal{H} = \prod_k \mathcal{H}_k \otimes \mathcal{H}_{-k}$, where $\mathcal{H}_k$ and $\mathcal{H}_{-k}$ are the Hilbert spaces of the modes of the particle and the antiparticle, respectively.
The full `in' vacuum state $\ket{0}_{\text{in}}$ is described by 
\eq{
	\ket{0}_{\text{in}}
&=
	\prod_{k,-k}
	\ket{0_k }_{\text{in}}
	\otimes
	\ket{0_{-k} }_{\text{in}}
\equiv
	\prod_{k,-k}
	\ket{0_k 0_{-k}}_{\text{in}}
,
}
where
\eq{
	a^{\text{in}}_k
	\ket{0_k }_{\text{in}}
&=
	b^{\text{in}}_{-k}
	\ket{0_{-k} }_{\text{in}}
=
	0
,
\label{eq:defofvacuum}
}
and likewise for the `out' states.
The  state $\ket{0_k 0_{-k}}_{\text{in}}$ can be expanded in terms of the `out' states as 
\eq{
	\ket{0_k 0_{-k}}_{\text{in}}
&=
	\sum_{n=0}^\infty
	C_{n_k}^0
	\ket{n_k n_{-k}}_{\text{out}}	
,
\label{eq:relabetinandout}
}
by using the Schmidt decomposition.
The normalisation, ${}_{\text{in}} \angleN{0_k 0_{-k}|0_k 0_{-k}}_{\text{in}}=1$, yields 
$\sum_{n=0}^\infty \absN{C_{n_k}^0}^2=1.$

The properties of $C_{n_k}^0$ and the Bogoliubov transformation (\ref{eq:Bogoliuboftrf}) yield the recurrence relation $ C_{n_k}^0 = \roundN{\beta_k / \alpha_k} C_{(n-1)_k}^0$, giving,
$
	C_{n_k}^0
=
	\roundN{
	{\beta_k}/{\alpha_k}
	}^n
	C_{0_k}^0
$
as discussed in~\cite{Ebadi:2014ufa}.
Using now this relation and (\ref{eq:defofalphabeta}),
we obtain
\eq{
	\absLR{C_{0_k}^0}
&=
	\frac{
	1
	}{
	\absLR{\alpha_k}
	}
=
	\frac{
	1
	}{
	\sqrt{2\pi}
	}
	\absLR{
	\Gamma(-\nu)
	e^{i \pi(1+\nu)/2}
	}
.
\label{eq:absC00}
}
Then we derive
$
    C_{n_k}^0
=
	\sqrt{1-\absLR{\gamma}^2}
	\gamma^n
	e^{i\theta_\nu^0}
$,
    $\gamma
=
	\squareN{1+e^{\pi \mu}}^{-1/2}
	\exp \squareLR{i \roundLR{ {3\pi}/{4} + \arg{\Gamma(-\nu)}}}
$ and
$
    \theta_\nu^0
=
	{\pi}/{4}
	+
	\arg{\Gamma(-\nu)}
	+
	\phi_c
$,
where $\phi_c$ is a constant.
Note that $\absN{\gamma} < 1/\sqrt{2}$ when $\mu > 0$, and hence $C_{n_k}^0$ approaches $0$ as the label $n$ increases.

Let us comment on other features of $C_{n_k}^0$.
First, $C_{n_k}^0$ depends on only the variable $\mu$, Eq.~(\ref{mu}). Thus 
$C_{n_k}^0$ reflects the charge and the mass but not the momentum $k^0$ and $k^x$ as the feature of the (anti)particle.
Second, when $\mu \to \infty$, $ \absN{ C_{n_k}^0 } \to \delta_{n0}$ since $\absN{\gamma}$ approaches $0$, and hence (\ref{eq:relabetinandout}) becomes $
	\ket{0_k 0_{-k}}_{\text{in}}
\to
	C_{0_k}^0
	\ket{0_k 0_{-k}}_{\text{out}}	
$, where the difference between the left- and right-hand side is just the phase factor.
Third, when $\mu \to 0$, $\absN{\gamma}$ approaches $1/\sqrt{2}$, and hence $\absN{ C_{n_k}^0 } \to 2^{-(n+1)/2}$.

The density matrix for the `in' vacuum state $\ket{0_k 0_{-k}}_{\text{in}}$
 is given by 
$
	\rho^{(\text{v})}
=
	\ket{0_k 0_{-k}}_{\text{in}}
	\hspace{-0.7ex}
	\bra{0_k 0_{-k}}
$, which is a pure state.
Employing (\ref{eq:relabetinandout}), we obtain the reduced density matrix for the particle as
$
	\rho_k
=
	\Tr_{-k}
	\rho^{(\text{v})}
=
	\sum_{n=0}^\infty
	\absN{C_{n_k}^{0}}^2
	\ket{n_k}_{\text{out}}
	\hspace{-0.6ex}
	\bra{n_k}
$,
and hence the entanglement entropy, defined in Section \ref{sec:EE},
is give by 
\eq{
	S_k
&=
	-
	\Tr_k
	\rho_k
	\log \rho_k
=
	-
	\absLR{\beta_{k}}^2
	\log \absLR{\beta_{k}}^2
	+
	\roundLR{
	1
	+
	\absLR{\beta_{k}}^2
	}
	\log
	\roundLR{
	1
	+
	\absLR{\beta_{k}}^2
	}
,
\label{eq:EEforrhovk}
}
where $
    {}_{\text{in}}\braN{0_k} 	
    a^{\text{out}}_k{}^\dagger
    a^{\text{out}}_k
	\ketN{0_k }_{\text{in}}
=
    {}_{\text{in}}\braN{0_{-k}} 	
    b^{\text{out}}_{-k}{}^\dagger
    b^{\text{out}}_{-k}
	\ketN{0_{-k} }_{\text{in}}
= 
    \absN{\beta_{k}}^2 
=
    \exp\squareN{-\pi \mu}
$, is the density of created particles.
We are dealing with a pure state, and hence 
$S_k=S_{-k}$.

\begin{figure}
	\centering
		\includegraphics[scale=.55]{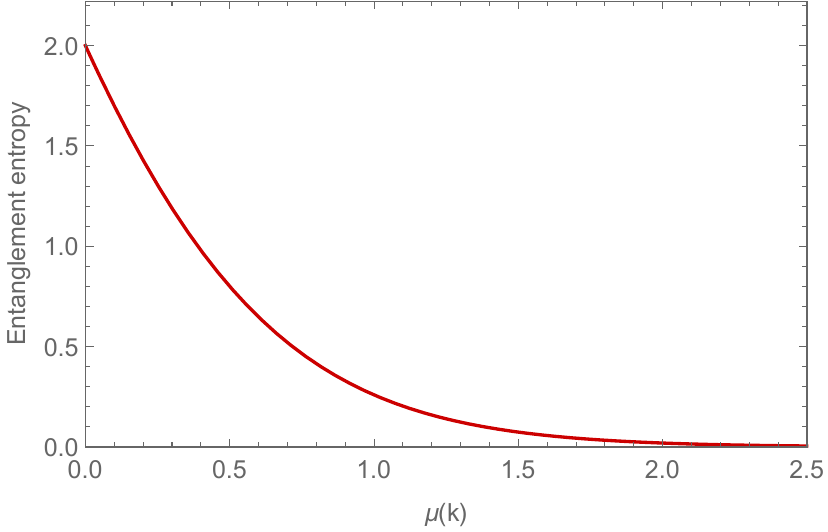}
		\caption{
		The entanglement entropy $S_k$ of the vacuum state $\ket{0_k 0_{-k}}$ vs. the parameter $\mu$ given by Eq.~(\ref{mu}).
		Considering $m^2$ and $E$ to be fixed,  $S_k$ is maximum in the small $\mu$ ($\absN{B}, n_L \ll \absN{E}$, along with our initial condition, $m^2/|qE|\ll1$), where the number of outgoing particle is unity, and vanishing for large $\mu$ ($E \ll n_L, \absN{B}$), where the number density of outgoing particle is zero. See the text for discussions.
		}
		\label{fig:EEforVacuum1}
\end{figure}

We obtain the $\mu$-dependence of $S_k$ as shown in Fig.\ref{fig:EEforVacuum1}.
Thus $S_k$ decreases as $\mu$ increases, and it is maximum, $S_k=2$, in the limit $\mu \to 0$, where $\absN{\beta_k}^2$ becomes unity. On the other hand, $S_k \to 0$ in the limit $\mu \to \infty$, where the reduced density matrix $\rho_k$ returns to the incoming pure state and $\absN{\beta_k}^2$ becomes zero.
This corresponds to the suppression of pair creation due to the stabilisation of the vacuum with increasing $B$.

\section{Mutual information and logarithmic negativity in systems of two scalar fields}
\label{sec:bipartite}
\noindent
Let us now consider systems which are constructed by two complex scalar fields.
There are two species of (anti)particles, which do not interact with each other.
The total Hilbert space $\mathcal{H}$ is given by, $\mathcal{H} = \prod_{s,k} \mathcal{H}_s \otimes \mathcal{H}_{-s} \otimes \mathcal{H}_k \otimes \mathcal{H}_{-k}$, where $ s$ and $ k$ stand for the two species of scalar fields. We assume that these two scalar fields have the same charge, but different masses are allowed.

We shall focus on the maximally entangled states for the incoming states of the (anti)particles. Now, the gauge transformation properties of the wave function of a charged field in quantum electrodynamics  puts a constraint on how one can prepare those states, as follows. The wave functions corresponding to two states with different charge content will have different transformation properties under the local gauge transformation. Hence if we add two or more states to construct an entangled state, we must ensure that the charge content of each of these states are the  {\it same}, so that the wave function for the full state has a definite transformation property. This will be reflected in the states (\ref{charge1}) and (\ref{eq:PAstate})
we work with.

\subsection{Single-charge state}
\label{sec:1particle}
\noindent
Based upon the above argument, we consider a maximally entangled single-charge state,
$
	\rho^{(1)}
=
	\ketN{\psi_{sk}^{(1)}}
	\braN{\psi_{sk}^{(1)}}
$, which is a pure state, with
\eq{
	\ket{\psi_{sk}^{(1)}}
&=
	\frac{\ket{0_s 0_{-s}; 1_k 0_{-k}}_{\text{in}}+\ket{1_s 0_{-s}; 0_k 0_{-k}}_{\text{in}}}{\sqrt{2}}
\label{charge1}
}
In our notation, the first (second) pair of entries appearing in the kets stands for the first (second) scalar. For a specific pair, the first (second) entry represents particle (antiparticle).  

Using the expansion of the incoming vacuum, (\ref{eq:relabetinandout}), we rewrite $
	\ket{ 1_k 0_{-k}}_{\text{in}}
=
	{ a^{\text{in}}_k }^\dagger
	\ket{ 0_k 0_{-k}}_{\text{in}}
$
by the outgoing states as
\eq{
	\ket{ 1_k 0_{-k}}_{\text{in}}
&=
	\sum_{n=0}^\infty
	C_{n_k}^1
	\ket{(n+1)_k n_{-k}}_{\text{out}}	
,
\label{eq:10byout}
}
where the coefficient $C_{n_k}^1$ is given by
\eq{
	C_{n_k}^1
=
	\frac{\sqrt{n+1}}{\alpha_k}
	C_{n_k}^0
=
	\roundLR{1-\absLR{\gamma_k}^2}
	\gamma_k^n
	e^{i\theta_\nu^1}		
,
\qquad \qquad 
	\theta_\nu^1 
=
	2 
	\roundLR{
	\frac{\pi}{4} 
	+ 
	\arg{\Gamma(-\nu)}) 
	}
	+ 
	\phi_c
\label{eq:C1nkandtheta1}
}
Here, we are using the label $k$ for $\gamma_k$, as it depends on the mass and the charge of the particle with  momentum $k$.
The features of $C_{n_k}^1$ are given in parallel with that of  $C_{n_k}^0$ in the preceding section.

The coefficients $C_{n_k}^1$ depend on only the variable $\mu$, and hence $C_{n_k}^1$ reflects the charge and the mass but not the momentum $k^0$ and $k^x$.
When $\mu \to \infty$, we obtain $ \absN{ C_{n_k}^1 } \to \delta_{n0}$, and hence (\ref{eq:10byout}) becomes $
	\ket{ 1_k 0_{-k}}_{\text{in}}
\to
	C_{0_k}^1
	\ket{1_k 0_{-k}}_{\text{out}}	
$, where the difference between the left- and right-hand side is just a phase factor.
When $\mu \to 0$, we have $\absN{ C_{n_k}^1 } \to \sqrt{n+1}/ 2^{(n+2)/2}$.

Also, using the relations (\ref{eq:relabetinandout}) and (\ref{eq:10byout}), the single-particle `in' state $\ketN{\psi_{sk}^{(1)}}$ can be written in terms of the `out' states, necessary to make the squeezed state expansion.

\subsubsection{Quantum Mutual information}
\noindent
Here we compute the quantum mutual information defined in Section \ref{sec:QMI},
corresponding to the state in Eq.~(\ref{charge1}). We shall focus on two reduced density matrices that characterise the particle-particle and also the particle-antiparticle correlations between the two scalar fields.

Let us start with the particle-particle correlation.
The reduced density matrix is given by $
	\rho_{s,k}^{(1)}
=
	\Tr_{-s,-k}
	\rho^{(1)}
$ and is written in terms of the `out' states as 
\eq{
	\rho_{s,k}^{(1)}
&=
	\frac{1}{2}
	\sum_{n,\ell=0}^\infty
	\roundB{
	C_{\ell_s}^0
	C_{n_k}^1
	\ket{\ell_s (n+1)_k}_{\text{out}}
	+
	C_{\ell_s}^1
	C_{n_k}^0
	\ket{(\ell+1)_s n_k}_{\text{out}}
	}
	\times
	\roundLR{
	\text{h.c.}
	}
,
\label{eq:rho1sk}
}
where $\roundN{\text{h.c.}}$ stands for the Hermitian conjugate of the first parenthesis.
In the limit of large $\mu(k)$ and $\mu(s)$,
$S(\rho_{s,k}^{(1)})$ vanishes.

\begin{figure}
	\centering
		\includegraphics[scale=.55]{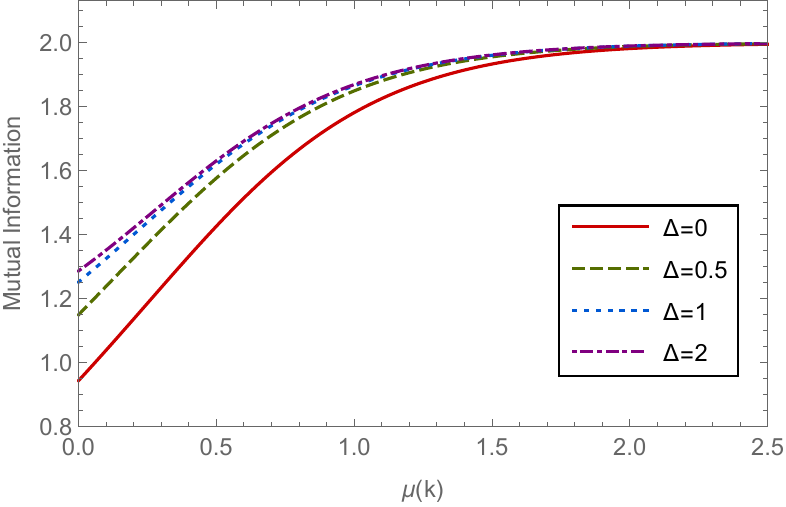}
		\caption{The quantum mutual information of $\rho_{s,k}^{(1)}$ vs. $\mu(k)$ (i.e., the particle-particle sector) for each value of $\Delta$, corresponding to the single charge state in Eq.~(\ref{charge1}). All lines approach $S(\rho_{s}^{(1)})+S(\rho_{k}^{(1)})=2$ as $\mu(k)$ increases. This asympotic value corresponds to the initial state itself, see text for discussion.}
		\label{fig:MIForRhoskForOP1}
\end{figure}

The quantum mutual information is defined by $
	I(\rho_{s,k}^{(1)})
=
	S(\rho_{s}^{(1)})
	+
	S(\rho_{k}^{(1)})
	-
	S(\rho_{s,k}^{(1)})
$,
where
$
	\rho_{s}^{(1)}
=
	\Tr_{k}
	\rho_{sk}^{(1)}
$ and $
	\rho_{k}^{(1)}
=
	\Tr_{s}
	\rho_{sk}^{(1)}
$.
The summation in (\ref{eq:rho1sk}) converges rapidly and hence for numerical purpose, we replace the infinity with a finite but large $n$- and $\ell$-value.
We thus obtain the $\mu$-dependence of $I(\rho_{s,k}^{(1)})$, shown in Fig.~\ref{fig:MIForRhoskForOP1}.
Here we have defined 
 $$\Delta \equiv \mu(s)-\mu(k),$$
reflecting, e.g., the mass difference between the fields. 
Moreover as we consider various parametric values of $\Delta$, we assume to vary $\mu(s)$ keeping $\mu(k)$ fixed.

Fig.~\ref{fig:MIForRhoskForOP1} shows that $I(\rho_{s,k}^{(1)})$ approaches its maximum value, $S(\rho_{s}^{(1)})+S(\rho_{k}^{(1)})=2$, as $\mu(k)$ increases, showing
the correlation of the particle-particle sector is maximum
for the large $\mu$ limit of (\ref{eq:rho1sk}).
When $\mu$ is small, the lines for the different values of $\Delta$ split, e.g., the mass difference of the two scalar fields can be estimated with fixed $E$ and $B$ in that region. 
Fig.~\ref{fig:MIForRhoskForOP1} also implies that the pair-creation disturbs the correlation, which originally exists in terms of the incoming modes.
\\

Next, we consider correlations of the particle-antiparticle and antiparticle-antiparticle sector.
We write the reduced density matrices,
$
	\rho_{\pm s,-k}^{(1)}
=	
	\Tr_{\mp s,k}
	\rho^{(1)}
$, in terms of  the outgoing modes as
\eq{
	\rho_{s,-k}^{(1)}
&=
	\frac{1}{2}
	\sum_{n,\ell=0}^\infty
	\roundB{
	C_{\ell_s}^0
	C_{(n-1)_k}^1
	\ket{\ell_s (n-1)_{-k}}_{\text{out}}
	+
	C_{\ell_s}^1
	C_{n_k}^0
	\ket{(\ell+1)_s n_{-k}}_{\text{out}}
	}
\times	\roundLR{
	\text{h.c.}
	}
,
\label{eq:rho1s-k}
}
\eq{
\rho_{-s,-k}^{(1)} 
&=
	\frac{1}{2}
	\sum_{n,\ell=0}^\infty
	\roundB{
	C_{\ell_s}^0
	C_{(n-1)_k}^1
	\ket{\ell_{-s} (n-1)_{-k}}_{\text{out}}
	+
	C_{(\ell-1)_s}^1
	C_{n_k}^0
	\ket{(\ell-1)_{-s} n_{-k}}_{\text{out}}
	}
\times	\roundLR{
	\text{h.c.}
	}
,
\label{eq:rho1-s-k}
}
with the requirement $C^1_{(-1)_k}=0$.
We also define 
$
	\rho_{-k}^{(1)}
=
	\Tr_{s}
	\rho_{s,-k}^{(1)}
$ and 
$
	\rho_{-s}^{(1)}
=
	\Tr_{-k}
	\rho_{-s,-k}^{(1)}
$.
Note that $\rho_{\pm s,-k}^{(1)}$ becomes a product state
$
	\rho_{\pm s}^{(1)}
	\otimes
	\rho_{-k}^{(1)}
$ 
in the limit of large $\mu(k)$ and $\mu(s)$, and consequently the mutual information becomes zero, as discussed in Section.~\ref{sec:QMI}.

\begin{figure}
	\centering
		\includegraphics[scale=.65]{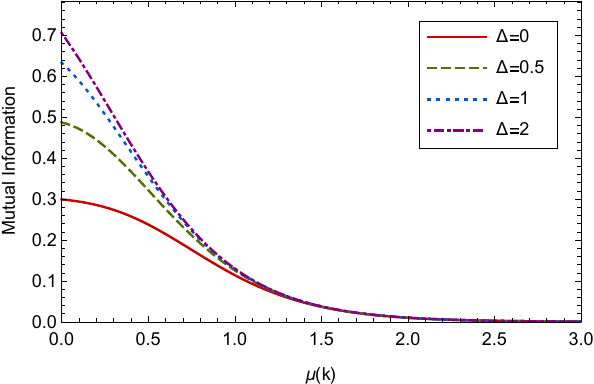}
		\includegraphics[scale=.5]{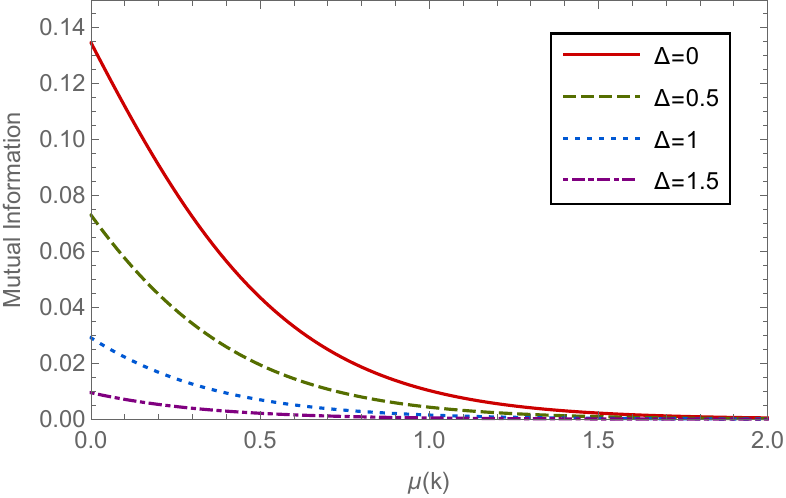}
		\caption{The quantum mutual informations of $\rho_{s,-k}^{(1)}$ (left) and of $\rho_{-s,-k}^{(1)}$ (right) vs. $\mu(k)$ for each value of $\Delta$ (i.e., the particle-antiparticle and antiparticle-antiparticle sector), corresponding to the single charge state in Eq.~(\ref{charge1}). 
		All lines approach zero, since $S(\rho_{\pm s,-k}^{(1)})=S(\rho_{\pm s}^{(1)})+S(\rho_{-k}^{(1)})$ in the limit of large $\mu(k)$. Note the	qualitative difference from Fig.~\ref{fig:MIForRhoskForOP1}. }
		\label{fig:MIForRhosmkForOP1}
\end{figure}

Fig.~\ref{fig:MIForRhosmkForOP1} shows that the mutual information of $\rho_{\pm s,-k}^{(1)}$ approaches zero as $\mu(k)$ increases. This corresponds to the fact,  that for large $\mu(k)$ values, the Bogoliubov transformation becomes trivial, and the `out' and `in' states coincide modulo some trivial phase factors, as discussed in Section~\ref{sec:1particle}. However, (\ref{charge1}) has no antiparticle content in it, resulting in a vanishing mutual information between the particle-antiparticle (antiparticle-antiparticle) sector in this limit.
On the other hand, for smaller $\mu$ values, the lines split as Fig.~\ref{fig:MIForRhoskForOP1}.
In addition, we observe that $\rho_{- s,-k}^{(1)}$ shows the inverted hierarchy of $\Delta$ compared to $\rho_{s,\pm k}^{(1)}$.

\subsubsection{Logarithmic negativity}
\begin{figure}
	\centering
		\includegraphics[scale=.65]{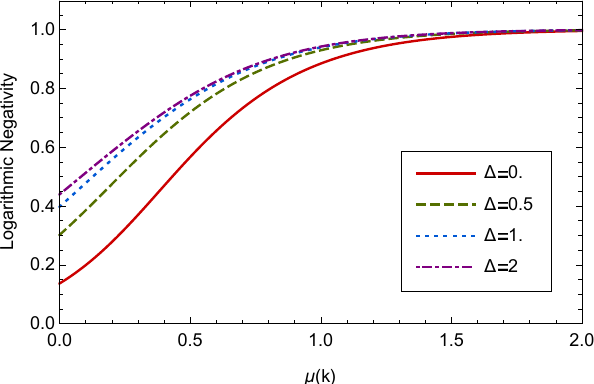}
		\caption{
		The logarithmic negativities of $\rho_{s,k}^{(1)}$ vs. $\mu(k)$  (i.e. the particle-particle sector) for each value of $\Delta$, corresponding to the single charge state in Eq.~(\ref{charge1}). In the limit of large $\mu(k)$, the logarithmic negativities approach unity. Also, we have not plotted the particle-antiparticle sector, where the logarithmic negativity turns out to be vanishingly small for all $\mu(k)$ values.
		}
		\label{fig:LNPskmu50}
\end{figure}

\noindent
Let us now compute the logarithmic negativity,
first for the particle-particle sector, $\rho_{s,k}^{(1)}$.
The $\mu(k)$-dependence of the logarithmic negativity of $\rho_{s,k}^{(1)}$ is shown in 
Fig.~\ref{fig:LNPskmu50} for different values of $\Delta$. The logarithmic negativity increases as $\mu(k)$ increases and for large $\mu(k)$-values, all the lines converge to unity.
This is because $\rho_{s,k}^{(1)}$ has the same eigenvalues as that of incoming modes in the large $\mu(k)$ limit, so that $\mathcal{L}_N \to \log_2 \left(4 \times {1}/{2}\right) = 1$.
This behaviour implies that the entanglement of $\rho_{s,k}^{(1)}$ is disturbed by the pair-creation.
For the particle-antiparticle and antiparticle-antiparticle sector, however, 
we find that the logarithmic negativities are vanishingly small, $\mathcal{L}_N(\rho_{\pm s, -k}^{(1)}) \lesssim{\cal O}(10^{-15})$, for all $\mu(k)$ values, showing once again the qualitative differences of these sectors with the particle-particle sector.

We shall consider  another example of entangled state below and will see the differences between the information quantities associated with it and those of Eq. (\ref{charge1}).

\subsection{Zero-charge state}
\label{sec:zerocharge}
\noindent
Keeping in mind the discussion made at the end of Section~\ref{sec:bipartite},
we now consider a pure state $
	\rho^{(\text{PA})}
=
	\ketN{\psi_{sk}^{(\text{PA})}}
	\braN{\psi_{sk}^{(\text{PA})}}
$, where 
\eq{
	\ket{\psi_{sk}^{(\text{PA})}}
&=
	\frac{
	1
	}{
	\sqrt{2}
	}
	\roundB{
	\ket{1_s 0_{-s}; 0_k 1_{-k}}_{\text{in}} 
	+ 
	\ket{0_s 1_{-s}; 1_{k} 0_{-k}}_{\text{in}} 
	},
\label{eq:PAstate}
}
has zero net charge content. Let us compute the same measures as earlier in order to see the qualitative differences.
\subsubsection{Quantum Mutual information}
\noindent
Using the earlier techniques, we derive the reduced density operator for the particle-particle sector,
$
	\rho_{s,k}^{(\text{PA})}
=
	\Tr_{-s,-k}
	\rho^{(\text{PA})}
$, which is written in terms of the `out' states as 
\eq{
	\rho_{s,k}^{(\text{PA})}
&=
	\frac{1}{2}
	\sum_{\ell,n=0}^\infty
	\roundB{
	C_{\ell_s}^1
	C_{(n-1)_k}^1
	\ket{(\ell+1)_s (n-1)_k}_{\text{out}}
	+
	C_{(\ell-1)_s}^1
	C_{n_k}^1
	\ket{(\ell-1)_s (n+1)_k}_{\text{out}}
	}
	\times
	\roundLR{
	\text{h.c.}
	}
.
\label{eq:rhoPAsk}
}
\begin{figure}
	\centering
		\includegraphics[scale=.5]{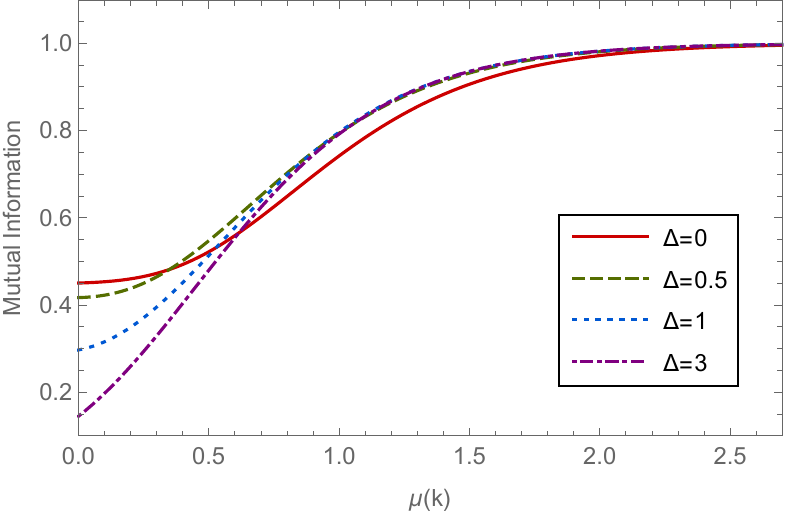}
		\caption{The quantum mutual information of $\rho_{s,k}^{(\text{PA})}$ vs. $\mu(k)$ (i.e., the particle-particle sector) for each value of $\Delta$, corresponding to the particle-antiparticle state in Eq.~(\ref{eq:PAstate}).}
		\label{fig:MIForRhoskForPA}
\end{figure}
On the other hand, the reduced density operator for the  particle-antiparticle sector is given by
\eq{
	\rho_{s,-k}^{(\text{PA})}
&=
	\frac{1}{2}
	\sum_{n,\ell=0}^\infty
	\roundB{
	C_{\ell_s}^1
	C_{n_k}^1
	\ket{(\ell+1)_s (n+1)_{-k}}_{\text{out}}
	+
	C_{(\ell-1)_s}^1
	C_{(n-1)_k}^1
	\ket{(\ell-1)_s (n-1)_{-k}}_{\text{out}}
	}
	\times
	\roundLR{
	\text{h.c.}
	}
.
\label{eq:rhoPAs-k}
}
\begin{figure}
	\centering
		\includegraphics[scale=.5]{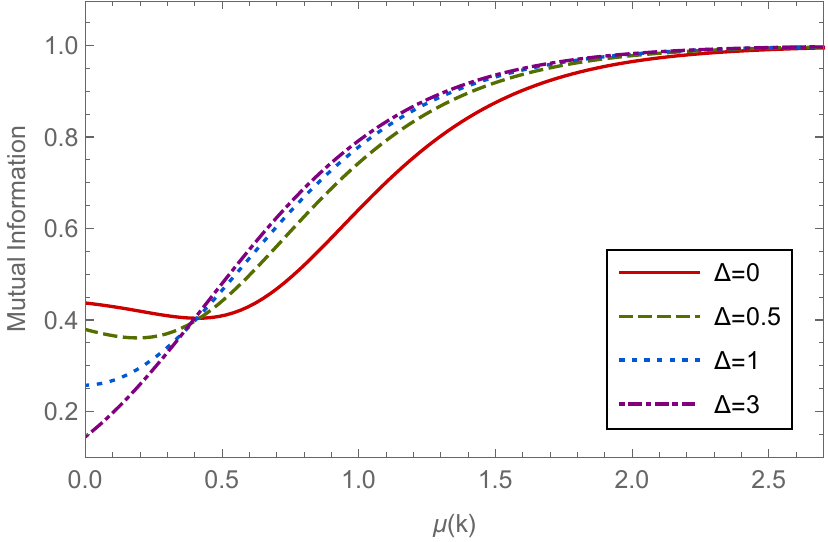}
		\caption{The quantum mutual information of $\rho_{s,-k}^{(\text{PA})}$ vs. $\mu(k)$ (i.e., the particle-antiparticle sector) for each value of $\Delta$, corresponding to the particle-antiparticle state in Eq.~(\ref{eq:PAstate}).}
		\label{fig:MIForRhos-kForPA}
\end{figure}
The $\mu(k)$-dependence of the quantum mutual information corresponding to the particle-particle sector, Eq.~(\ref{eq:rhoPAsk}), and the particle-antiparticle sector, Eq.~(\ref{eq:rhoPAs-k}), for different values of $\Delta$ is depicted respectively in Fig.~\ref{fig:MIForRhoskForPA}, Fig.~\ref{fig:MIForRhos-kForPA}. We note the overall qualitative similarity between them, although Fig.~\ref{fig:MIForRhos-kForPA} shows minimum for $\Delta \to 0$. 
We also note that the mutual information for both of them have the same numerical orders, unlike those of the single-charge state.
These similarities should correspond to the symmetry in particle and antiparticle number of the initial state, Eq.~(\ref{eq:PAstate}). Due to the same reason, the antiparticle-antiparticle sector of this state yields exactly the same correlations as those of particle-particle sector, and we do not pursue it any further.

We also note some qualitative difference in the behaviour of Fig.~\ref{fig:MIForRhoskForOP1} with that of Fig.~\ref{fig:MIForRhoskForPA} and  Fig.~\ref{fig:MIForRhos-kForPA}:
the former has simple feature such as the monotonicity and unchanging hierarchy with respect to $\Delta$, while the latter shows interchanges of the hierarchy and the non-monotonic behaviour in Fig.~\ref{fig:MIForRhos-kForPA}.

\subsubsection{Logarithmic negativity}
\begin{figure}
	\centering
		\includegraphics[scale=.5]{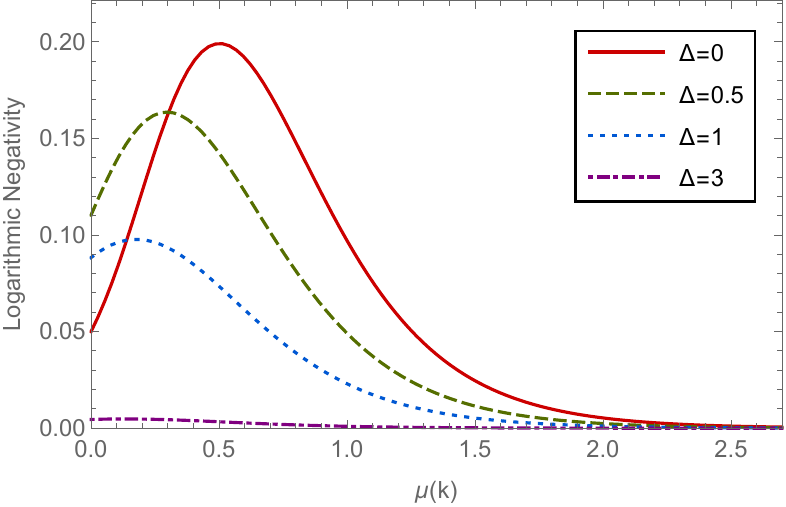}
		\caption{The logarithmic negativity of $\rho_{s,k}^{(\text{PA})}$ vs. $\mu(k)$ (i.e., the particle-particle sector) for each value of $\Delta$, corresponding to the particle-antiparticle state in Eq.~(\ref{eq:PAstate}).}
		\label{fig:LNForRhoskForPA}
\end{figure}

\begin{figure}
	\centering
		\includegraphics[scale=.5]{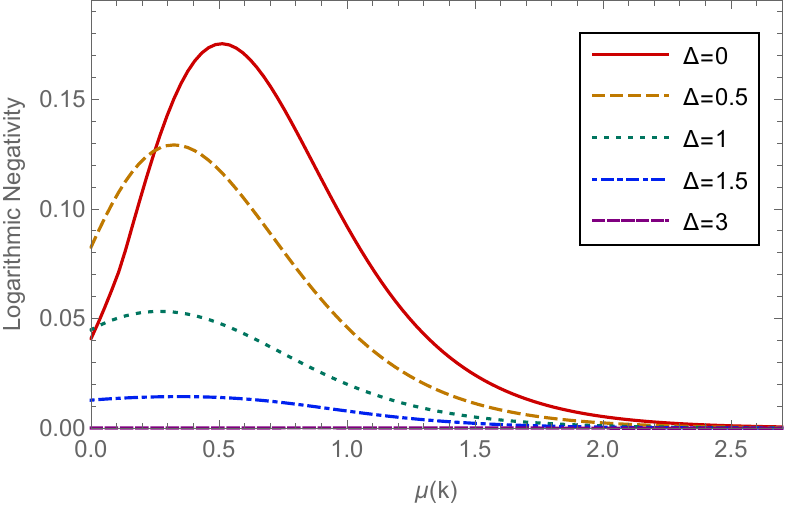}
		\caption{The logarithmic negativity of $\rho_{s,-k}^{(\text{PA})}$ vs. $\mu(k)$ (i.e., the particle-antiparticle sector) for each value of $\Delta$, corresponding to the particle-antiparticle state in Eq.~(\ref{eq:PAstate}).}
		\label{fig:LNForRhos-kForPA}
\end{figure}

 Finally, we compute the logarithmic negativity for the particle-particle and the particle-antiparticle sector by following the methods described earlier. They have respectively been plotted in  Fig.~\ref{fig:LNForRhoskForPA} and Fig.~\ref{fig:LNForRhos-kForPA}. We note the overall qualitative similarity between them, owing once again to the symmetry in the number of particles and antiparticles in (\ref{eq:PAstate}). The asymptotically vanishing values in the plots for large $\mu(k)$  just as earlier correspond to the fact that in this limit we reach the initial state, Eq.~(\ref{eq:PAstate}), and it  has no logarithmic negativity, as can be checked easily. 
The antiparticle-antiparticle sector behaves in the same manner as the particle-particle sector.
We also note the qualitative differences of these plots with that of Fig.~\ref{fig:LNPskmu50}, which arises from the existence of both particle and antiparticle in the incoming zero-charge state.

\noindent

\section{Summary and outlook}
\label{sec:SD}
\noindent
We now summarise our results. The chief motivation of this work was to quantify the effect of a background magnetic field on the quantum correlations between the Schwinger pairs. We have studied the vacuum entanglement entropy in Section~\ref{sec:vacuum}, and the quantum mutual information and logarithmic negativity for maximally entangled  states with  single and zero electric charges respectively in Section~\ref{sec:1particle} and Section~\ref{sec:zerocharge}. We have emphasised the qualitative differences in the behaviour of the information quantities between these  states. 
Note also that since the number density of created particles equals $|\beta_k|^2=e^{-\pi \mu}$ (cf. the discussion below  Eq.~(\ref{eq:EEforrhovk})), all the plots above will show similar behaviour with respect to $|\beta_k|^2$ as well. 
Extension of these results to the Rindler and the inflationary backgrounds seems to be interesting.

Finally, we note that in all the plots the various information quantities converge to some specific points for sufficiently large $\mu(k)$ values. Assuming for example, the mass, the electric field and the Landau level to be fixed, a large $\mu(k)$ corresponds to large values of the magnetic field, Eq.~(\ref{mu}).
In this limit  the Bogoliubov transformation becomes trivial  and an `out' state becomes coincident with the `in' state, modulo some trivial phase factor (cf., the discussion below Eq.~(\ref{eq:absC00})).
Keeping in mind that the background electric field is analogous to the acceleration parameter in an accelerated frame as far as the particle creation is concerned, it then seems possible that the degraded quantum correlation between two entangled states in such frames, e.g.~\cite{MartinMartinez:2010ar}, might be restored (for charged fields) via the application of a background magnetic field, as follows. The magnetic Lorentz force, $q \vec{v}\times \vec{B} $, acts in the same direction for the particle and antiparticle initially moving in the opposite direction just after the pair creation. An electric field or background spacetime curvature/acceleration do just the opposite effect by moving the created pairs away, e.g.~\cite{Mironov:2011hp}. Thus to the best of our understanding, it seems reasonable to expect that the particle-antiparticle thermal pair creation causing the entanglement degradation in the Rindler frame, will get diminished in the presence of a background magnetic field of an `appropriate' value. As an example, we may consider Fig.~\ref{fig:LNForRhos-kForPA}. The logarithmic negativity, which is a measure of entanglement for a mixed ensemble indeed grows and reaches maximum for certain $\mu$ values. After this point however, the particle creation becomes too weak and the squeezed state expansion coincides with that of the initial state, which itself has vanishing logarithmic negativity. Depending upon the characteristics of the initial state, analogous arguments can be made for all the cases we have investigated in this paper.   We hope to come back to this issue in detail in our future work. Such effect can in particular be relevant for a black hole endowed with a strong magnetic field in its exterior.

\section*{Acknowledgments}
\noindent
SB is partially and  HH is fully supported by the ISIRD grant 9-289/2017/IITRPR/704.
SC is partially supported by the ISIRD grant 9-252/2016/IITRPR/708.

\end{document}